\pdfoutput=1

\documentclass[prd, aps, twocolumn, nofootinbib,superscriptaddress]
{revtex4}

\usepackage{amsmath}
\usepackage{amssymb}
\usepackage{latexsym}
\usepackage{graphicx,subfigure,graphics,epsfig,amsmath,amssymb}
\usepackage{hyperref}
\usepackage{bm}
\usepackage{color}

\newcommand{\be}{\begin{equation}}
\newcommand{\ee}{\end{equation}}
\newcommand{\beqa}{\begin{eqnarray}}
\newcommand{\eeqa}{\end{eqnarray}}

\newcommand{\AS}{A_{\rm S}}
\newcommand{\AT}{A_{\rm T}}
\newcommand{\nS}{n_{\rm S}}
\newcommand{\nT}{n_{\rm T}}

\newcommand{\impc}{\ensuremath{{\rm\,Mpc}^{-1}}}

\newcommand{\simlt}{\lower.5ex\hbox{$\; \buildrel < \over \sim \;$}}
\newcommand{\simgt}{\lower.5ex\hbox{$\; \buildrel > \over \sim \;$}}

\newcommand{\iMpc}{\impc}

\usepackage{natbib}

\begin{document}

\title{Tensors, BICEP2, prior dependence, and dust}
\author{Marina Cort\^{e}s} 
\affiliation{Institute for Astronomy, University of Edinburgh, Royal Observatory, Edinburgh EH9 3HJ, United Kingdom}
\affiliation{Perimeter Institute for Theoretical Physics 31 Caroline Street North, Waterloo, Ontario N2J 2Y5, Canada} 
\author{Andrew R. Liddle}
\affiliation{Institute for Astronomy, University of Edinburgh, Royal Observatory, Edinburgh EH9 3HJ, United Kingdom}
\author{David Parkinson}
\affiliation{School of Mathematics and Physics, University of Queensland,
  Brisbane, QLD 4072, Australia} 
\date{\today}

\begin{abstract} 
We investigate the prior dependence on the inferred spectrum of primordial tensor perturbations, in light of recent results from BICEP2 and taking into account a possible dust contribution to polarized anisotropies. We highlight an optimized parameterization of the tensor power spectrum, and adoption of a logarithmic prior on its amplitude $\AT$, leading to results that transform more evenly under change of pivot scale. In the absence of foregrounds the tension between the results of BICEP2 and {\it Planck} drives the tensor spectral index $\nT$ to be blue-tilted in a joint analysis, which would be in contradiction to the standard inflation prediction ($\nT<0$). When foregrounds are accounted for, the BICEP2 results no longer require non-standard inflationary parameter regions. We present limits on primordial $A_{\rm T}$ and $n_{\rm T}$, adopting foreground scenarios put forward by Mortonson \& Seljak and motivated by {\it Planck} 353~GHz observations, and assess what dust contribution leaves a detectable cosmological signal. We find that if there is sufficient dust for the signal to be compatible with standard inflation, then the primordial signal is too weak to be robustly detected by BICEP2 if {\it Planck}+WMAP upper limits from temperature and $E$-mode polarization are correct. 

\end{abstract}

\maketitle

\section{Introduction}

The announcement of detection of large-angle primordial B-mode polarization in the cosmic microwave background (CMB) by the BICEP2 experiment \cite{BICEP2} earlier this year caused considerable stir in the cosmology community, due to the possibility of the signal being due to gravitational waves. Primordial gravitational waves are almost exclusively a signature of the inflationary mechanism. The detection was not marginal, the headline value for the tensor-to-scalar ratio being $r=0.20^{+0.07}_{-0.05}$ with the null result disfavoured at $7$-sigma. These results incorporated polarized foreground mapping and characterization. Foreground estimates put forward by the team at the time represented a maximum of 20\% signal contamination, the most pessimistic foreground model, DDM2, bringing down $r$ to $r=0.16$. 

Subsequently, suspicions have grown that the contribution from dust foregrounds is larger than originally thought, and the published version of the BICEP2 paper notes that existing data cannot exclude the possibility of the observed signal being entirely due to such foregrounds~\cite{BICEP2}. Studies by Mortonson and Seljak \cite{Mortonson:2014} and Flauger \textit{et al.}~\cite{Flauger:2014} used preliminary maps from the {\it Planck} satellite and inferred a template for polarized dust contamination which, extrapolated to the BICEP2 patch, could potentially completely account for the B-mode signal detected by the BICEP2 team. The Planck collaboration has now released results \cite{planckdust} showing that this high dust amplitude is indeed the most likely outcome of extrapolation from their 353~GHz channel maps, though the uncertainty remains broad. On a more optimistic note, Colley and Gott \cite{colleygott} conclude using genus topology that the imperfect match between {\it Planck} $Q$ and $U$ Stokes' parameter maps and the BICEP2 maps implies roughly half the observed signal cannot be attributed to dust.

In the early stages after the detection, a focus of the community was on the apparent discrepancy between BICEP2's detection and {\it Planck}'s upper bound on $r$ of 0.11 at 95\% confidence \cite{planck} (though see Ref.~\cite{Audren:2014} for a discussion of how real the discrepancy actually is given the different scales probed by the experiments). There were various attempts at addressing the discrepancy by invoking a cosmological origin. These branched mainly into investigating modifications of the scalar sector of the perturbations as well as parameters which are degenerate with it \cite{MHA,AAEP,ADSS,Li:2014cka,Smith:2014kka,WLLQCZ,Martin:2014lra}, 
and into analyses considering a positive value for the tilt of the tensor perturbations $n_{\rm T}$ 
\cite{WX1,Li:2014cka,Cheng:2014bma,Gerbino:2014,Chang:2014loa,Hu:2014aua,CW1,MNPS} which would be in contradiction to normal models of inflation. However, these analyses make prior assumptions in the modelling of the tensor perturbations which we shall show may be inappropriate.

In this article, our first aim is to establish a set of principles for defining the prior space of models including tensors, building on our earlier paper on the prior dependence of tensor constraints \cite{Cortes:2011}. This is the topic of the next section. Having set this framework, we then first revisit the analysis of tensor spectrum constraints under the assumption of the BICEP2 signal being entirely primordial, before extending the analysis to include models of dust contribution to the observed signal.

\section{Formulating prior assumptions}\label{s:prior}

In this section we lay down a set of principles for fixing prior assumptions for tensor mode data analysis. Progressively, they are as follows.
\begin{enumerate}
\item In an era where tensor detection is an objective, it is preferable to constrain the primordial tensor spectrum directly, rather than its ratio to the scalar spectrum.
\item As the order of magnitude of the tensor amplitude is {\it a priori} unknown, the prior distribution of tensor amplitudes must be chosen with care.
\item The tensor spectrum should be constrained at a `pivot' scale optimized for the set of data and model priors being considered.
\end{enumerate}

The BICEP2 detection prompted a number of analyses under different model assumptions. Typically the tensor-to-scalar ratio $r$ has been constrained, though Ref.~\cite{Chang:2014loa} considered the tensor amplitude directly. The pivot scale has normally been taken at a default value, such as the CosmoMC default of $0.05 \, {\rm Mpc}^{-1}$, or a different scale chosen but not optimized.

Concerning the prior distribution of tensor amplitudes, all articles to date have assumed a uniform prior on $r$ or on the tensor amplitude, even in cases where strongly blue-tilted spectra are considered \cite{prebigbang,ekp}. This is extremely hard to justify, as such a prior is uniform only at the chosen pivot scale and will be highly non-uniform at any other scale, as shown in Ref.~\cite{Cortes:2011}. Obviously there is no reason why the mechanism producing the perturbations should be aware of the scale at which we are able to constrain them, and have the special property of uniformity there. Hence it is crucial at least to test possible prior dependence of any conclusions being derived, and ideally to impose a more physically-motivated prior in the first place.

We now discuss these points in detail.

\subsection{The case for separate scalars and tensors} 

We parameterize our set of primordial spectra as simple power laws,
\beqa
\AS(k)&=&\AS(k_0) k^{\nS -1}\,,\\
\AT(k)&=&\AT(k_0) k^{\nT}\,,\label{AT}
\eeqa
where $k_0$ is the pivot scale where observables are specified at, and the spectral indices defined by
\beqa
\nS-1 \equiv \frac{d \ln \AS(k)}{d \ln k}\,, \quad \nT \equiv \frac{d \ln \AT(k)}{d \ln k}\,,
\eeqa
are taken to be constants throughout. 
The ratio of tensor-to-scalar amplitude of perturbations is defined as 
\be
r(k) \equiv \frac{\AT(k)}{\AS(k)} \,.\label{r}
\ee

Commonly the amplitude of B-modes is quantified by the fraction of tensor-to-scalar signal, $r(k)$, that could be constrained. 
This combination is well justified as long as we don't have a tensor amplitude detection, i.e.\ while the scalar perturbation is the only sector observed. If there is a tension between different limits on $r$ coming from different scales we can alleviate it by changing the shape of the scalar spectrum or by considering modifications to parameters that are degenerate with the scalar spectrum. 
However none of these modifications to $A_{\rm S}(k)$ help us to learn directly about the tensor sector, which is the main aim when we consider constraints on $r$.

For the case of BICEP2, proposals for reducing the tension with the bounds imposed by {\it Planck} include modifications to the running of the spectral index, spatial curvature, optical depth, effective number of neutrino species, etc.~\cite{Li:2014cka}.  Alleviating this tension with other data in this fashion is more a reflection of the way the scalar and tensor spectra are tied together and less of increased insight into the model behind the origin of fluctuations. We  argue that $r(k)$ is obsolete once there is a detection of primordial modes, which we want to characterize independently of the other parameters of the theory.
Tensions between datasets should be identified and accounted for on the basis of the parameters  appearing naturally in the underlying model.

\begin{figure} [t]
\includegraphics[width= 0.235\textwidth]{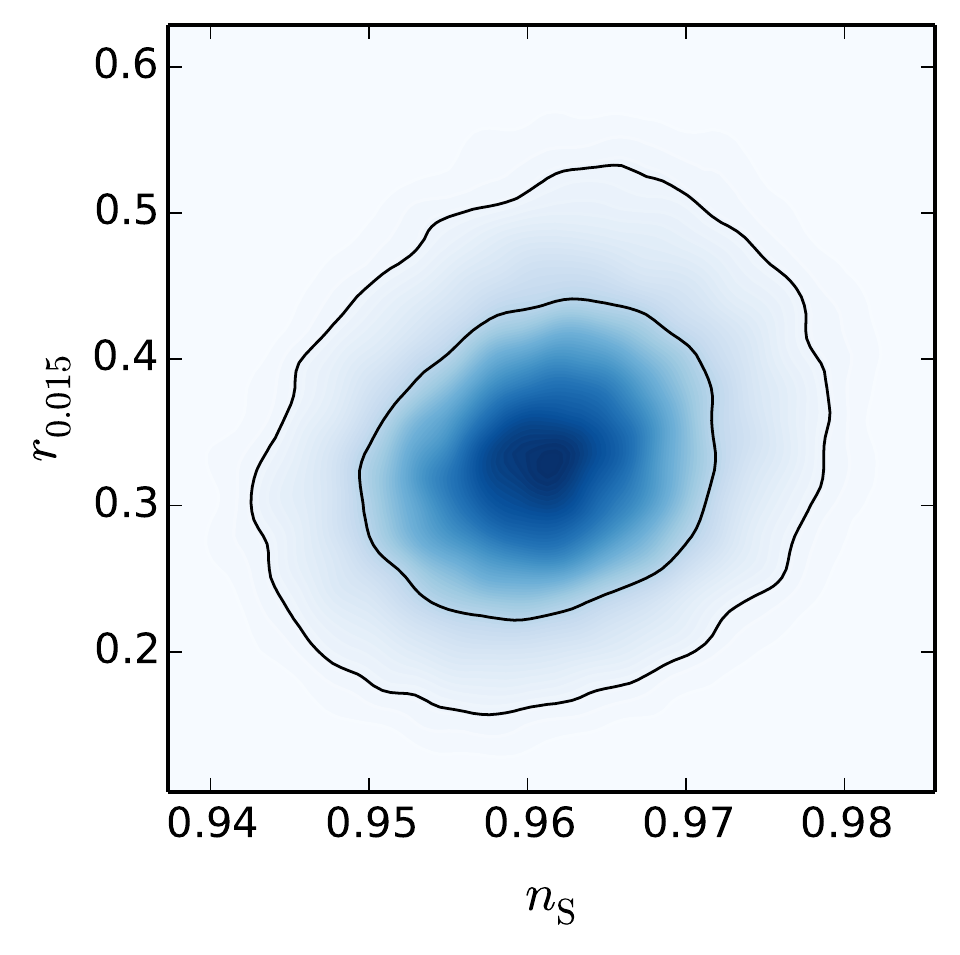}
\includegraphics[width= 0.235\textwidth]{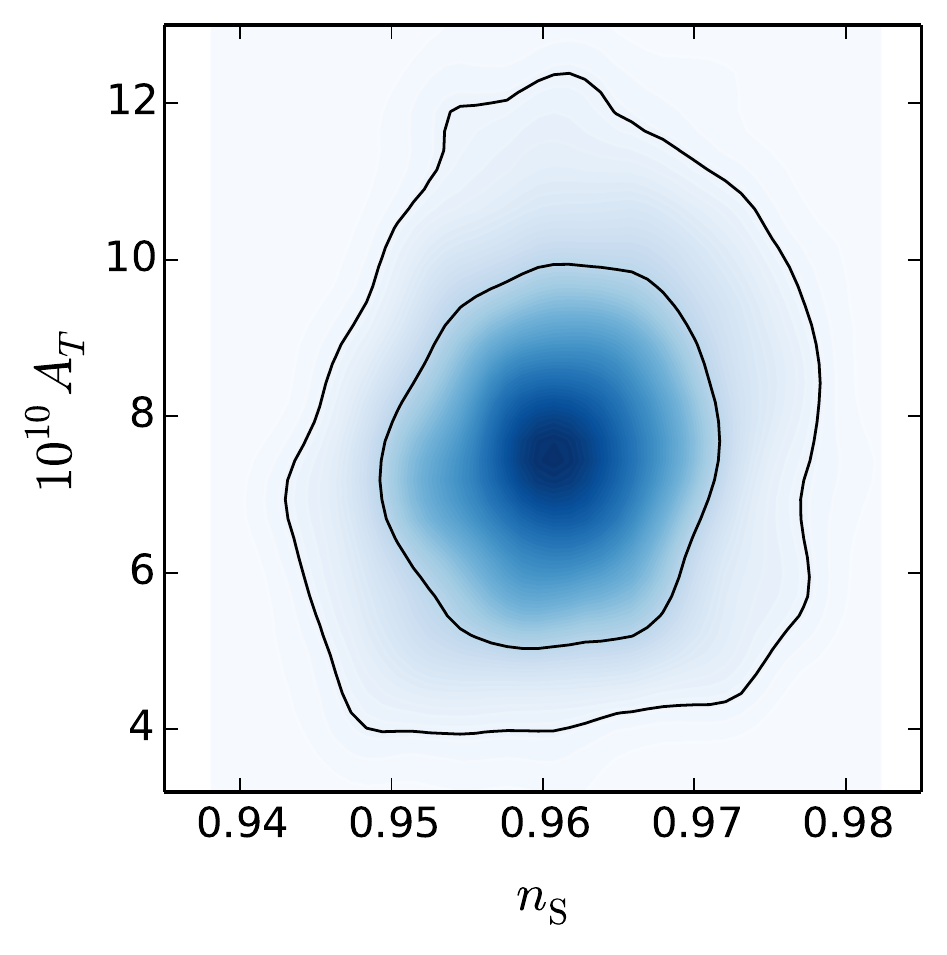}
\caption{Comparison between a fit to the tensor-to-scalar-ratio {\bf (left)}, and a fit the linear tensor amplitude $\AT$ {\bf (right)} both against the scalar tilt $\nS$, for a hypothetical case of no dust contribution to the B-mode signal and using the methods described below. Separating scalar from tensor variables, shown in the right panel, has the advantage of decorrelating the corresponding quantities.}
\label{tensorAmplitude} 
\end{figure}

While the above case for using $\AT$ is primarily theoretical, there is also benefit in reducing the correlation to scalar perturbation variables.
For example, in Fig.~\ref{tensorAmplitude} we show a comparison between fitting the amplitude of the tensor modes $\AT$ as opposed to the tensor-to-scalar ratio; the former shows a mild positive correlation while the latter shows none. The advantage of separating the scalars from the tensors in this example is modest. However we would expect that if the tensor detection was less significant, i.e.\ less than 7-sigma, the correlation between scalars and tensors would be larger and the gain of decorrelating both quantities would be more visible.
Absence of correlations means that the constraining power of the data can be summarized with less information as one-dimensional projections of the constraints contain all the information within the two-dimensional plot. 
\subsection{Linear versus logarithmic prior on the tensor amplitude}

All analyses to date that combine BICEP2 with other CMB data used a uniform prior on $\AT$ or $r$  
\cite{Li:2014cka, Cheng:2014bma,Gerbino:2014,Smith:2014kka}, with the exception of Ref.~\cite{Chang:2014loa}. There is no reason to apply a uniform prior on the scale at which an experiment measures $A_{\rm T}$, because no physical model will single out that one scale as the one to consider a prior to be uniform at, as opposed to any other scale. A safer prior is the Jeffreys' prior \cite{Jaynes}, which is typically applied when a positive-definite continuous quantity is analysed and whose order-of-magnitude is unknown, as is the case with $A_{\rm T}$.\footnote{We don't have complete uncertainty about the tensor spectrum. We know it is positive definite, and though we don't know the order of magnitude, we know it is driven by new physics somewhere between the electroweak scale and the GUT scale. Thus the ``order-of-order of magnitude'' is known.} This prior takes a logarithmic form which is justified by invariance under change of parameterization.

\begin{figure*} [t]
\begin{center}
\includegraphics[width= 0.4\textwidth]{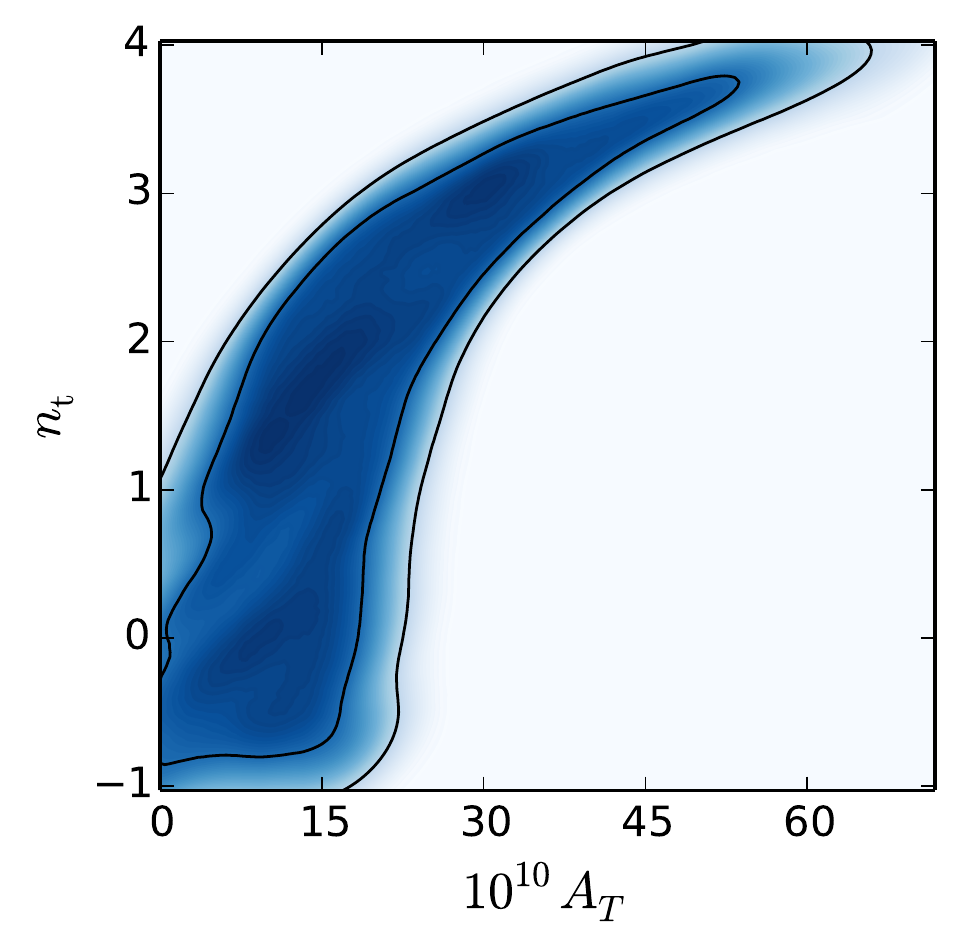}
\includegraphics[width= 0.4\textwidth]{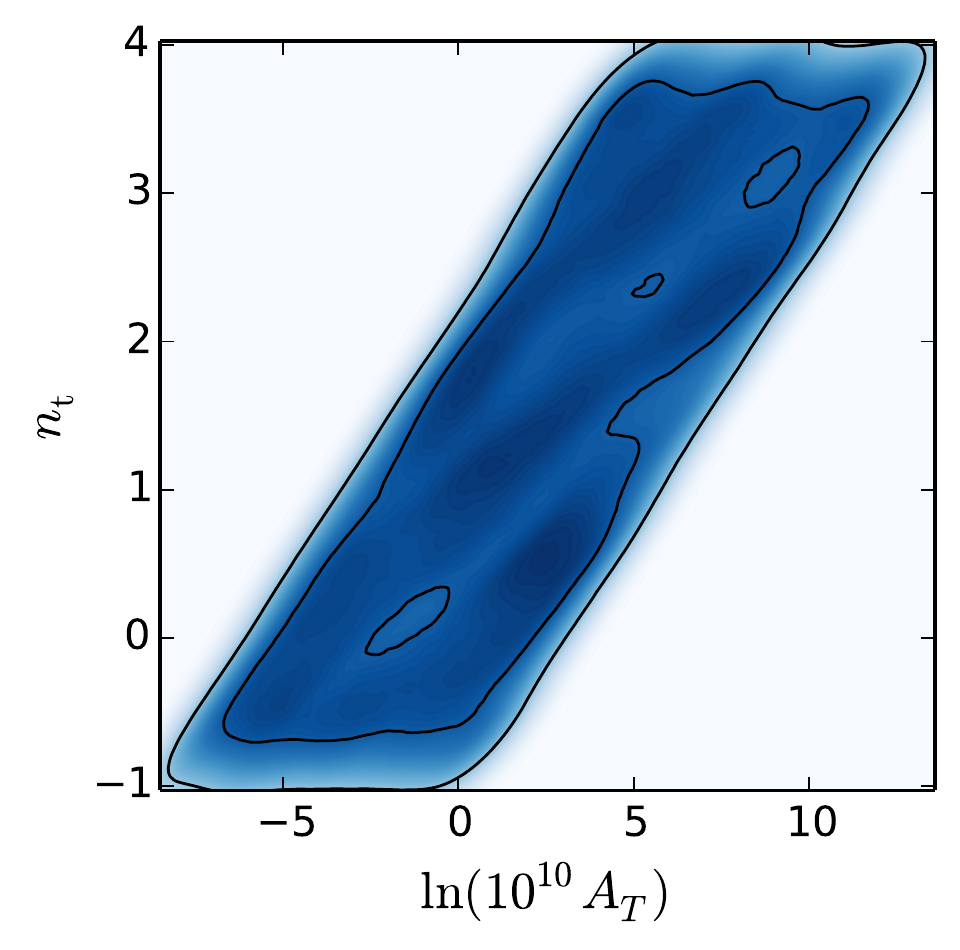}
\caption{Transformation of an uniform prior density with cosmological scale, from $k=0.001 \impc$ to $k=0.015 \impc$. {\bf Left panel}: Linear prior in $\AT$,  uniform density at the original scale does not correspond to uniform density at the transported scale, and we obtain distorted density contours at the new scale. {\bf Right panel}: Logarithmic prior on $\AT$ preserves the density of the contours between scales and hence ensures for safe transformation of the posterior between different $k$. We stress there is no data at all in both figures; we just draw uniform points on one scale and analytically transform r to the second scale ($\nT$ does not change).}
\label{priorTransform} 
\end{center}
\end{figure*}

%

Importantly, we will see in the next subsection that the logarithmic prior has well-behaved properties under change in pivot scale, as compared to the linear prior.
As we showed in Ref.~\cite{Cortes:2011}, a prior uniform on either $A_{\rm T}$ or $r$ doesn't correspond to a uniform prior at any other scale, because $\AT$ doesn't transform linearly with scale $k$. Its $k$-dependence, given by Eq.~(\ref{AT}), is exponential in $\nT$. 
In Fig.~\ref{priorTransform} (left panel) we show an example of the transformation of the prior on $\AT$ taken to be uniform at $k=0.002 \impc$ and transported to $k=0.01 \impc$. At the new scale the prior distribution is clearly not uniform. This means that in choosing to sample $A_{\rm T}$ uniformly at a given scale, we are singling out that scale as the only one where the prior is uniform, and all other scales are sampled non-uniformly. Priors uniform in $A_{\rm T}$ are not preserved under scale transformations. 

Instead, if we sample uniformly in $\ln \AT$ the transformation law is now linear in $\ln \AT$ and ensures preservation of the prior when transported across pivot scales.
The same is valid for $r$, with the added mixing of the joint transportation of the prior on both $\AT$ and $\AS$ (though for the latter the posterior is very well constrained within the prior so the same issues don't arise). In the right-hand panel of Fig.~\ref{priorTransform} we show the transformation of a prior uniform in $\ln \AT$, which apart from boundary effects remains uniform at the transformed scale. 

\subsection{The choice of pivot scale}

An advantage of separating the scalars from the tensors is the ready identification of a pivot scale for each corresponding to the experiment and observable we're constraining. In Refs.~\cite{Cortes:2007,Cortes:2011} we stressed the importance of choosing an optimized pivot scale for a parameter when quoting constraints on that parameter. We also noted the possibility of choosing separate pivot scales for the scalars and tensors, since even a given single experiment probes those most sensitively on different length scales.

The pivot scale of an observational dataset that measures tensor modes is the scale that decorrelates the uncertainties on $A_{\rm T}$ and its derivative $\nT$. This is different from the scale that decorrelates uncertainties on $r$ and its derivative, as this scale is also sensitive to the pivot scale for the scalar spectrum which is typically on shorter scales due to the different shape of the induced CMB power spectrum.

Since the BICEP2 release, there has been confusion in the literature as to what scale to choose for different datasets \cite{Li:2014cka, Cheng:2014bma,Gerbino:2014,Chang:2014loa}.
Some of this confusion was cleared up in Ref.~\cite{Audren:2014}, though again we point out that once the tensor contribution has been clearly detected, parameterization in terms of $r(k)$ is no longer necessary. 

In the following section we extract the pivot scales for the dataset combinations of interest.

\begin{figure*} [t]
\begin{center}
\includegraphics[width=\textwidth]{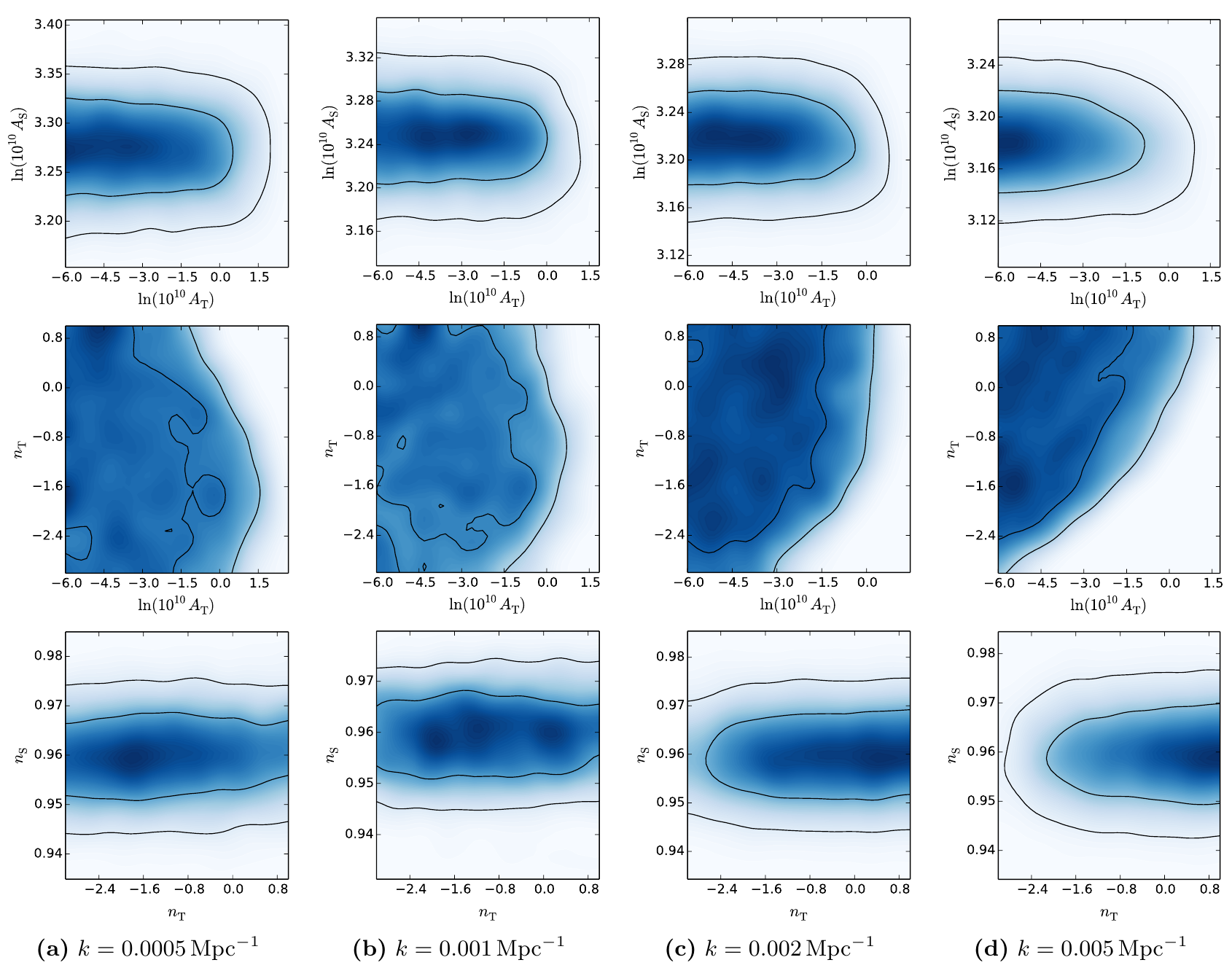}
\caption{Study for Planck's tensor pivot scale. The correlation between $\nT$ and $\ln \AT$ is minimized in {\bf (b)}, at $k=0.001 \impc$, which indicates the pivot scale. On scales away from the pivot a projection effect of the constraints gives evidence for a redder tilt if we probe at larger scales {\bf (a)}, and bluer tilt at smaller scales, {\bf (c)} and {\bf (d)}, a consequence of the prior.}
\label{planckpivot} 
\end{center}
\end{figure*}


\section{BICEP2 as a primordial signal}

We now derive constraints on the tensor spectrum using the optimal prior for each data combination. In this section we will assume that the BICEP2 signal is entirely primordial, so as to enable comparison with various previous works that have made different prior assumptions. The following section will incorporate models of polarized dust foregrounds.

First we identify appropriate scales for the combination of {\it Planck} temperature and WMAP polarization data, referred to as {\it Planck}+WP, and for the {\it Planck}+WP+BICEP2 combination. Starting with {\it Planck}+WP, we take the priors on the tensor parameters to be uniform in the ranges $-6 < \ln(10^{10} A_{\rm T}) < 3$ and $-3 < \nT <1$. The other cosmological parameters have the default priors set in the April 2014 CosmoMC release \cite{CosmoMC}, with foreground parameters handled as in the {\it Planck} collaboration analyses~\cite{planck}.

\begin{figure} [t]
\begin{center}
\includegraphics[width= 0.48 \textwidth]{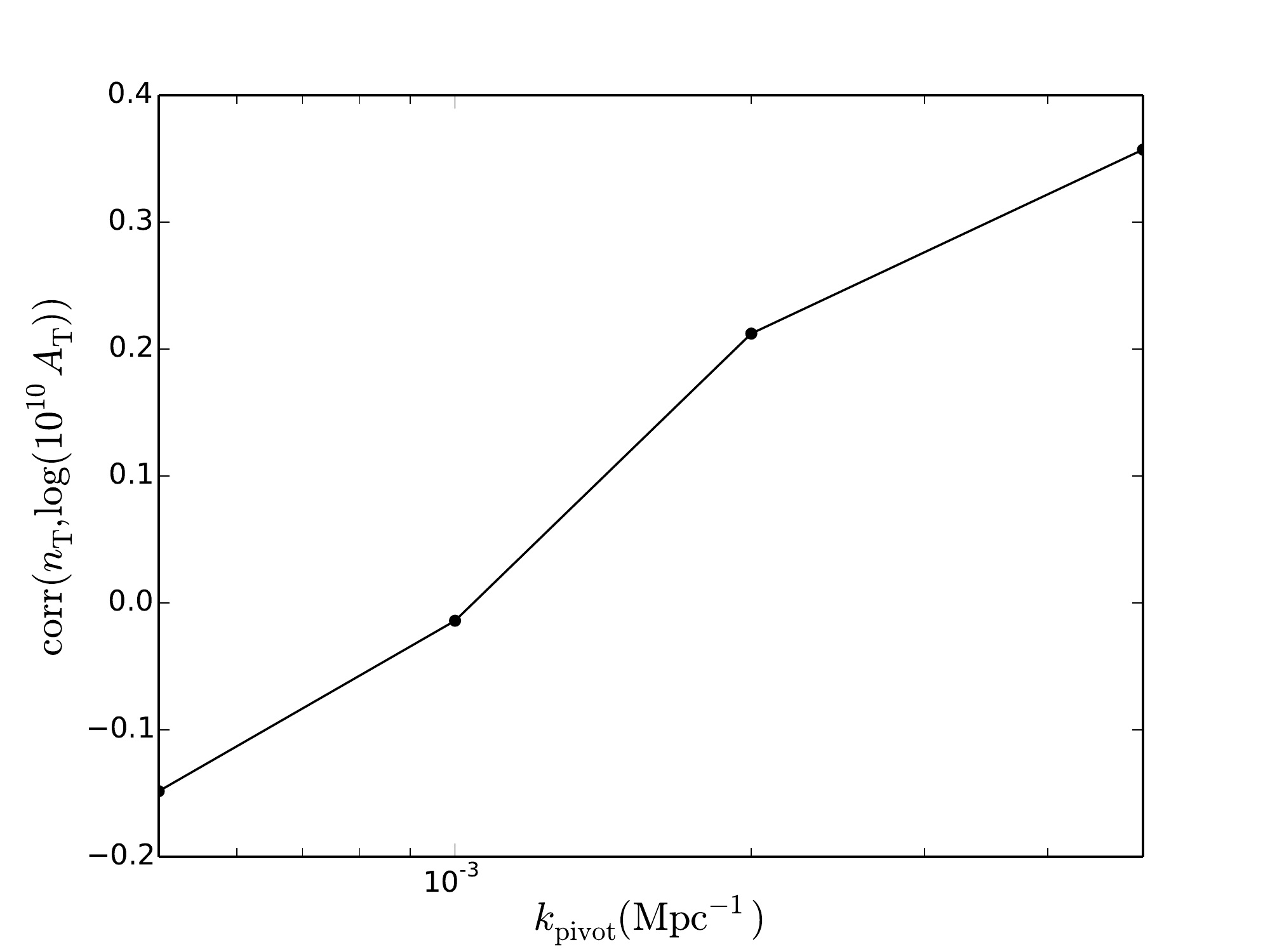}
\caption{The correlation coefficient at different pivots for {\it Planck}+WP. It crosses zero around $k = 0.001 \iMpc$.}
\label{f:corr} 
\end{center}
\end{figure}

On its own {\it Planck}+WP does not detect any tensor signal, but nevertheless the decorrelation technique of Ref.~\cite{Cortes:2007} can be used to estimate the pivot and its uncertainty. We perform runs at different scales, shown in Fig.~\ref{planckpivot}. {\it Planck}+WP has sensitivity to tensors only on a narrow range of scales and the constraints on the amplitude of tensors will be optimal around those scales, at which constraints will be broadly insensitive to the tilt of the spectrum.  Figure~\ref{planckpivot} shows constraints for various choices of pivot, and Fig.~\ref{f:corr} shows the correlation coefficient between the tensor amplitude and tilt at each scale. The scale best probed by {\it Planck}+WP is close to $k=0.001 \, {\rm Mpc}^{-1}$. At this scale the 95\% upper limit on $\ln (10^{10} A_{\rm T})$ is 0.1. This would correspond to a strong upper limit on $r$ (about 0.04), but this is not to be taken very seriously because the limit largely arises from the {\em prior} distribution containing mostly models whose $r$ value is below the {\it Planck} sensitivity, and hence is not arising from the data.

 If we probe away from the pivot, this will be reflected in measuring preferred values for the tensor tilt, which are either blue or red according to whether we're probing at smaller or larger scales than the pivot. These are not real detections of tilt of the spectrum but result from a projection effect of constraints at the pivot. This can be seen in Fig.~\ref{planckpivot} for scale $k=0.0005 \impc$ for which red values of $\nT$ are slightly preferred, and at scales \mbox{$k=0.002 \impc$} and $k=0.005 \impc$ which give indication of bluer values. 

\begin{figure} [t]
\includegraphics[width= 0.48 \textwidth]{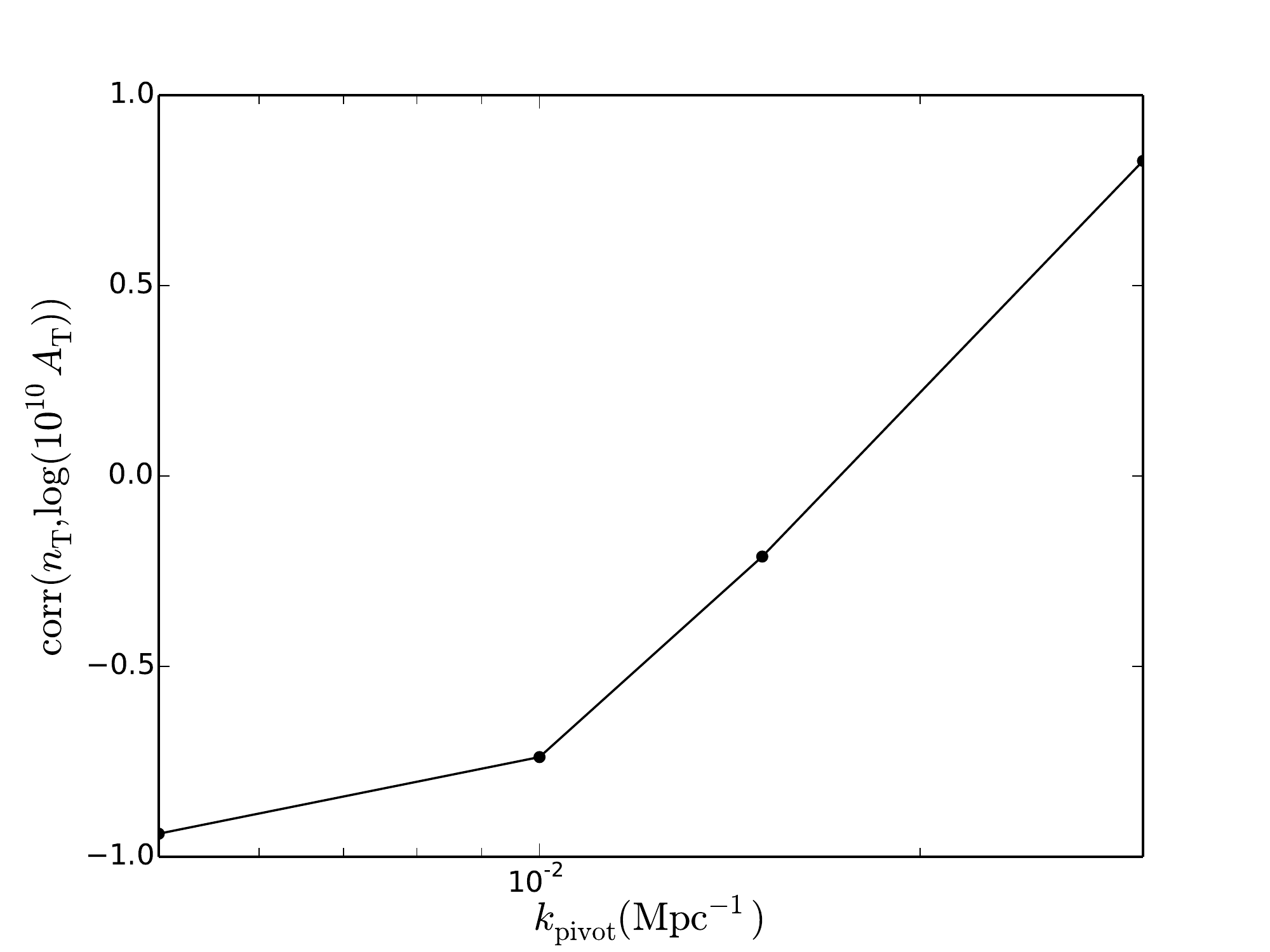}
\caption{As Fig.~\ref{f:corr}, for {\it Planck}+WP+BICEP2. It crosses zero around $k = 0.015 \iMpc$, in agreement with our extrapolation technique.}
\label{f:corrbicep} 
\end{figure}
\begin{figure} [t]
\begin{center}
\includegraphics[width= 0.3\textwidth]{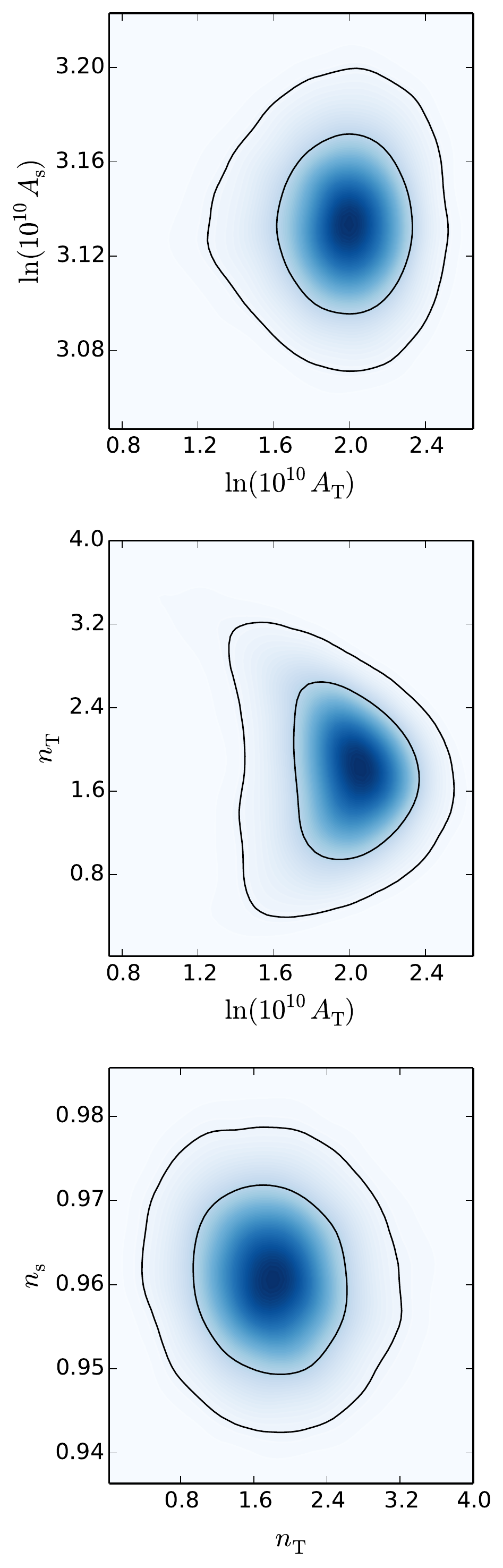}
\caption{Combined constraints from {\it Planck}+WP+BICEP2 at the decorrelation scale $k=0.015 \impc$.}
\label{bicep2pivot} 
\end{center}
\end{figure}

Adding BICEP2 data, when interpreted as wholly cosmological, gives a strong tensor detection. For this analysis we modify the prior on $\nT$ to the range \mbox{$-1 < \nT < 4$} in order to encompass the range that will be allowed by the data. We find that $k=0.015\pm0.002 \, {\rm Mpc}^{-1}$ corresponds to the scale best probed by {\it Planck}+WP+BICEP2 (roughly $\ell\sim150$). Indeed at this scale we find that $A_{\rm T}(k)$ is measured independently of $n_{\rm T}$, shown in Figs.~\ref{f:corrbicep} and \ref{bicep2pivot}. We have estimated the error on the pivot scale for {\it Planck}+WP+BICEP2 by considering runs at different scales, and extracting the corresponding pivots for each scale using the method described in Ref.~\cite{Cortes:2007}, which extrapolates to a scale that decorrelates a parameter and its derivative. The small difference between each of the pivots obtained in this way is indicative of the uncertainty in the value we adopt. 

For our main results in this section, the dataset combination of interest is {\it Planck}+WP+BICEP2, and the constraints are shown in Fig.~\ref{bicep2pivot}. We find $\ln (10^{10} A_{\rm T}) = 1.95_{-0.20}^{+0.27}$, corresponding to a central value $r = 0.32$. This exceeds the value quoted by BICEP2 because most of these models have $\nT > 0$ and the ratio is being quoted at a smaller scale. The significance of the detection is not nearly as strong as the uncertainty makes it appear (remember that the lower edge of our prior is at $\ln (10^{10} A_{\rm T}) = -6$, apparently a huge number of $\sigma$ away), because the likelihood does not fall further once the amplitude becomes too small to significantly affect the observables. The tensor spectral index is constrained as $n_{\rm T} = 1.8 \pm 0.6$. 

Our limits on $\nT$ are similar to those obtained by Gerbino et al.~\cite{Gerbino:2014}, who quote $\nT = 1.67 \pm 0.53$, though their fits did not vary other cosmological parameters and hence are not directly comparable. Chang and Xu quote the similar result $\nT = 1.70 \pm 0.57$ \cite{Chang:2014loa}. Much tighter constraints on $\nT$ with a lower central value consistent with zero, even just using BICEP2 data alone, were reported in Refs.~\cite{Cheng:2014bma}; we have not been able to understand why those results are so different from ours and others reported in the literature.

The strong preference for a blue-tilted spectrum is at odds with the prediction from single-field slow-roll inflation, $\nT=-2\epsilon$ where $\epsilon$ is the first slow-roll parameter $\epsilon(\phi)=\frac{1}{2} M_{\rm Pl}^2 (V'/V)^2$. Such blue-tilted tensor power spectra are predicted by inflation models that contain a `super-inflation' phase, for example those motived by Loop Quantum Gravity (e.g.\ Ref.~\cite{Copeland:2008kz}) as well as collapsing Universe models \cite{prebigbang,ekp}.

\section{Polarized foregrounds}

\subsection{Tensors in the presence of dust}

We now repeat the analysis of the previous section with the addition of candidate dust models based on the spectral shape of the polarized dust spectrum identified by {\it Planck} in regions of strong dust contribution. It has already been shown by Mortonson and Seljak \cite{Mortonson:2014} that if the dust amplitude is left as a free parameter, then it can readily soak up all of the large-angle B-mode signal, and then BICEP2's polarization is consistent with zero contribution from primordial modes. The {\it Planck} collaboration has shown that extrapolation from their 353~GHz observations indicates a dust contribution of this magnitude, though still with significant uncertainty \cite{planckdust}.

\begin{figure*} [t]
\begin{center}
\includegraphics[width= \textwidth]{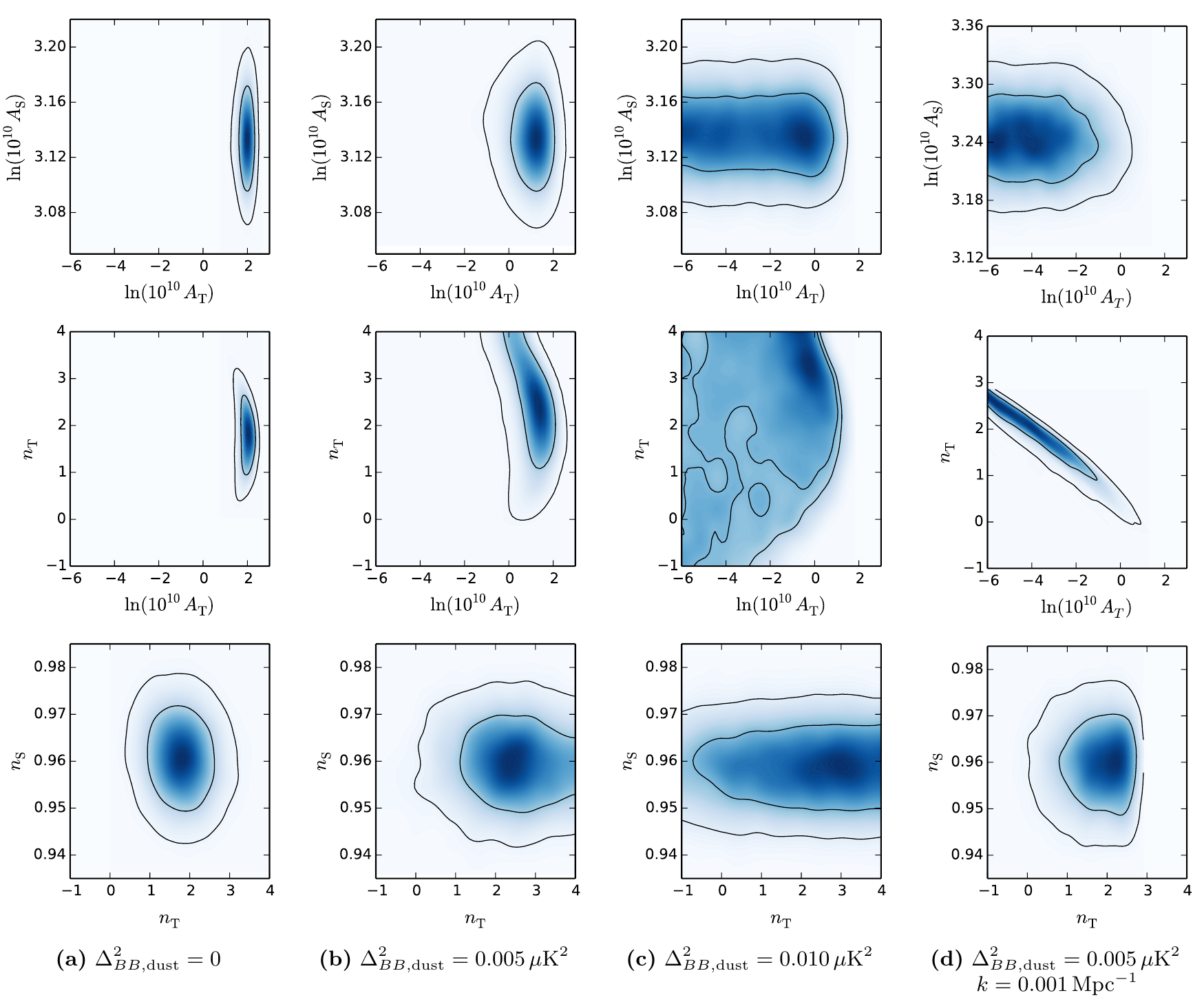}\\
\caption{Constraints on the tensor spectrum in the presence of foregrounds. {\bf (a)} has zero dust as in Fig.~\ref{bicep2pivot}.  {\bf (b)} has an `optimistic' value for the dust amplitude $\Delta^2_{BB,{\rm dust}}=0.005 \,\mu {\rm K}^2$, close to the lower limit in the analysis of Mortonson and Seljak, while {\bf (c)}  assumes a `pessimistic' value $\Delta^2_{BB,{\rm dust}}=0.010 \, \mu {\rm K}^2$, taken from the central value in the same analysis. {\bf (a)}, {\bf (b)} and {\bf (c)} are shown at the BICEP2's pivot $k=0.015\impc$. In {\bf (d)} we show constraints from the same dataset combination and the `optimistic' dust amplitude, but obtained at a different scale $k=0.001 \impc$. In this case we lose the detection of tensors.}
\label{dustOptimPessim} 
\end{center}
\end{figure*}


Rather than redo the analysis of Mortonson and Seljak, we envisage a future situation where the dust amplitude in the BICEP2 region has been accurately determined, and consider dust models with different but fixed overall amplitudes and spectral dependence. One point of exploration is whether inclusion of dust might permit negative $n_{\rm T}$, consistent with simple models of inflation, while still leaving a strong enough primordial signal to be detected. 

Mortonson and Seljak \cite{Mortonson:2014} expressed the dust contribution as a power law with fixed exponent, taking as free parameter the overall amplitude of the dust power, $\Delta^2_{BB,{\rm dust}}$ normalized at $\ell=100$.
Motivated by the values of the best-fit amplitude that they find, we carry out analyses for two possibilities for the dust component. One is for a pessimistic (i.e.\ large) value of the dust amplitude $\Delta^2_{BB,{\rm dust}}=0.010\, \mu {\rm K}^2$, corresponding to the best-fit of their analysis, and the other is for an optimistic value $\Delta^2_{BB,{\rm dust}}=0.005 \, \mu {\rm K}^2$ which is the lower $95\%$ confidence limit found in that work. We consider the same fixed spectral dependence $\Delta^2_{BB,{\rm dust}} \propto \ell^{-0.3}$. 

For comparison, the {\it Planck} collaboration report a dust power of 
\begin{equation}
{\cal D}^{BB}_\ell = 0.0132 \pm 0.0029 \; \mbox{(statistical)} \;  ^{+0.28}_{-0.24} \; \mbox{(systematic)}
\end{equation}
in a band centred on $\ell = 80$ \cite{planckdust}. Taking the liberty of adding the uncertainties in quadrature, as in their Fig.~9, and rescaling to $\ell = 100$ using either our adopted slope of $-0.3$ or their measured slope of $-0.42$, we find a 95\% confidence range for $\Delta^2_{BB,{\rm dust}}$ ranging from $0.005\, \mu {\rm K}^2$ to $0.020\, \mu {\rm K}^2$, i.e.\ the optimistic scenario we adopt is just allowed at 95\% confidence by {\it Planck}, while even our pessimistic scenario is below their best fit. On the positive side, our optimistic scenario is in good agreement with the result found by Colley and Gott using the genus statistic \cite{colleygott}. In any case, it is clear that current observations do not pin down the dust contribution at anything like the sensitivity that would be required to distinguish the scenarios that we are considering.

We show the results obtained in Fig.~\ref{dustOptimPessim}. The leftmost three columns adopt our standard pivot $k = 0.015 \iMpc$ and the logarithmic prior. The dust contribution increases from left to right. As dust increases, the inferred tensor amplitude reduces and the constraint on $\nT$ simultaneously weakens. For the optimistic (i.e.\ low) dust contribution model, the best-fit $\AT$ is reduced but there remains a detection at somewhat above 95\% confidence, while the allowed range for $\nT$ remains in the non-inflationary $\nT>0$ region. For the pessimistic dust model the detection is lost, to be replaced by an upper limit, and $\nT$ correspondingly becomes unconstrained. 

The outcome is that if {\it Planck}'s upper bounds on $r$ are correct, then inflation implies that BICEP2 cannot detect tensors at its sensitivity since they will be too small at BICEP2's scale. That means from the set \{BICEP2, {\it Planck}, Inflation models\} only two of these can be simultaneously consistent. The case that all three hold (here `BICEP2' meaning a detection of primordial tensors by that experiment) is not possible. 

\begin{figure} [t]
\begin{center}
\includegraphics[width=0.48 \textwidth]{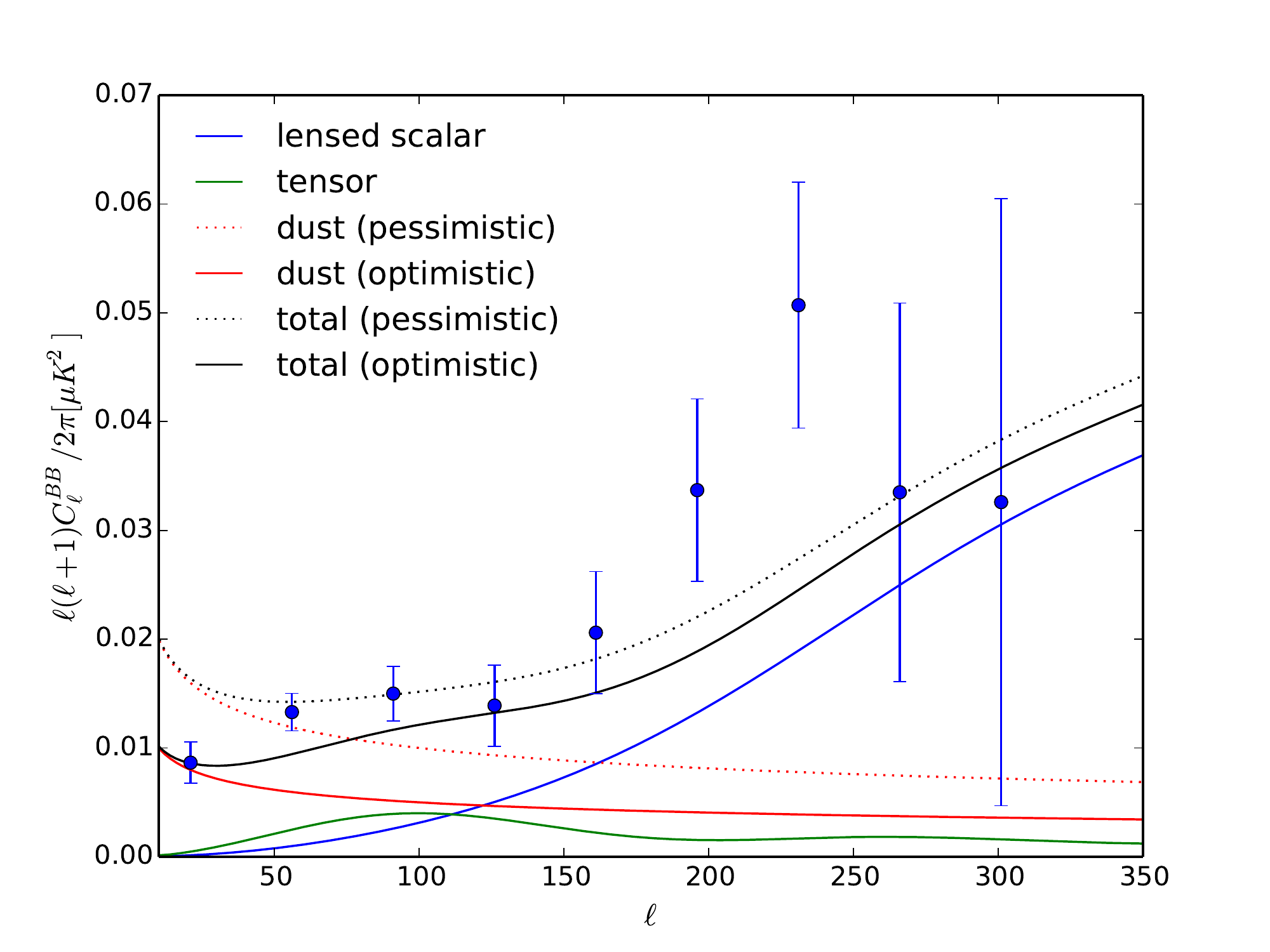}
\caption{The various contributions to the observed B-mode signal. The lensing contribution can be considered fixed, via the well-measured temperature anisotropies. We show the two different dust models considered in this article, optimistic and pessimistic. The green line indicates an example shape of the tensor spectrum, here with $\nT = 2.9$ (which is our best fit to the {\it Planck}+BICEP data), whose presence would be inferred if the sum of the other contributions falls short of explaining the full signal. Finally, the black lines show the totals obtained by summing lensing, this tensor shape, and each dust model.}
\label{compCls} 
\end{center}
\end{figure}

The value of dust amplitude we consider in Fig.~\ref{dustOptimPessim} corresponds to a fraction of foreground contribution to the overall B-mode signal of about 35\% (defined relative to the total $\ell(\ell+1)C_{\ell}^{BB}/2\pi$ evaluated at $\ell=46$). This can be used as a rule-of-thumb value, indicating the maximum contribution of the dust foreground that still preserves a primordial signal detection at 2-sigma at BICEP2 sensitivity.  The corresponding contributions to BB power from scalar lensing, primordial tensors, and foregrounds, in the pessimistic and optimistic scenarios, are shown in Fig.~\ref{compCls} together with BICEP2's band powers.

\begin{figure} [t!]
\includegraphics[width=0.49\textwidth]{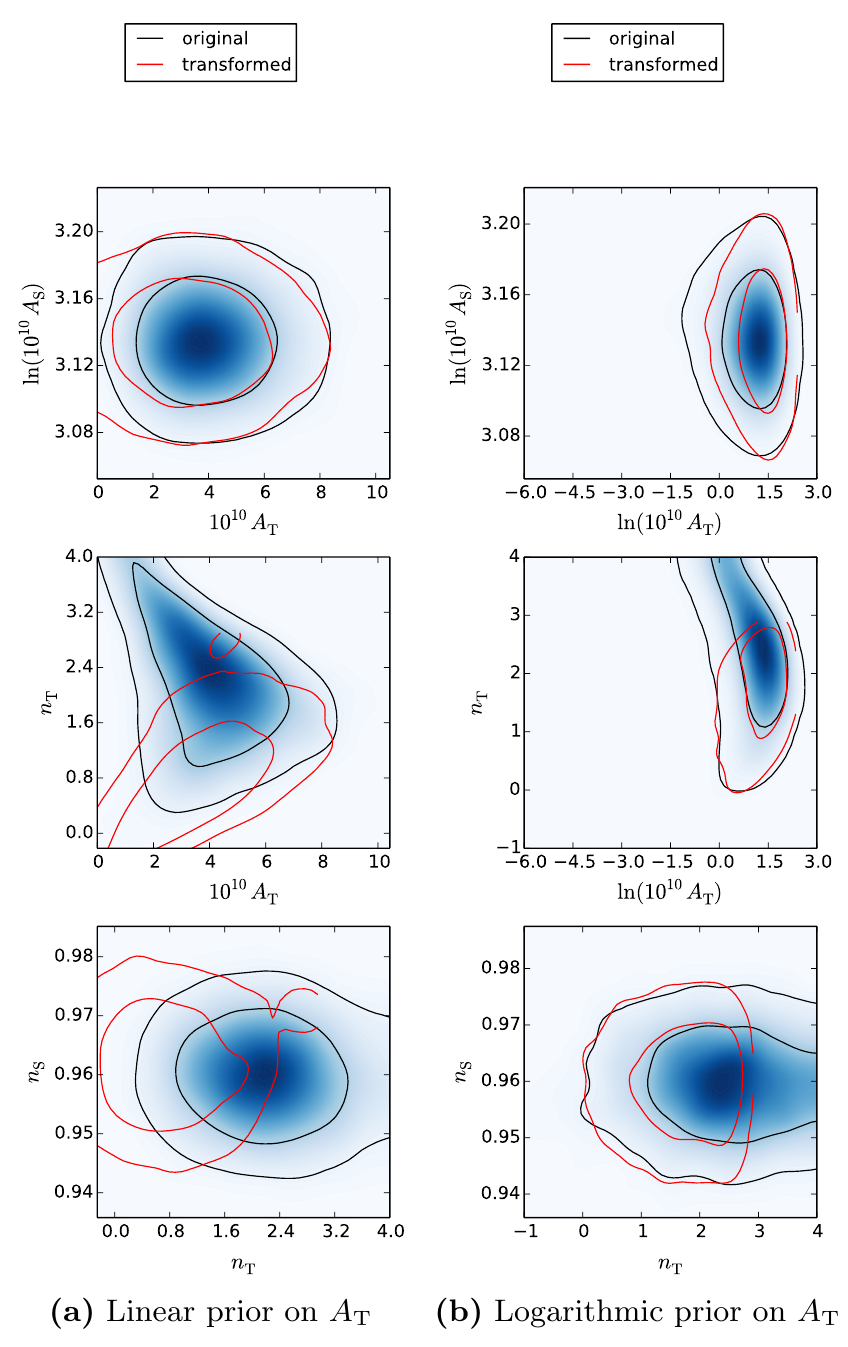}\\
\caption{Comparison of linear versus logarithmic prior on $\AT$ for the transformation between the scales of {\it Planck}+WP and {\it Planck}+WP+BICEP2, $k=0.001 \impc \rightarrow k=0.015\impc$.
\textbf{(a)} Linear prior on $\AT$: the blue contours are constraints obtained at the {\it Planck}+WP+BICEP2 pivot $k=0.015 \impc$ and the red contours are constraints obtained at the {\it Planck}+WP pivot scale, $k=0.001\impc$ and transported to the {\it Planck}+WP+BICEP2 scale \textbf{(b)} The same colour coding and scale transformation for a logarithmic prior on $\AT$. In this case the red contours show a mismatch in $\nT$ when transported to the new scale. Since the red contours are obtained at the Planck pivot, which is not the optimal scale, spectra which are very blue are removed by the lower limit on $\AT$. In the logarithmic case this appears as a simple cut at $n_\mathrm{T}=3$, while in the linear case it manifests as a shift in the probability density to lower values of $n_\mathrm{T}$.
These contours include a dust contribution of  $\Delta^2_{BB,{\rm dust}}=0.005\, \mu {\rm K}^2$. }
\label{posteriorTransform} 
\end{figure}


In the right-hand column of Fig.~\ref{dustOptimPessim} we show the constraints obtained for the optimistic dust amplitude $\Delta^2_{BB,{\rm dust}}=0.005 \,\mu {\rm K}^2$ as well as same remaining parameters, changing only the pivot scale to the {\it Planck} one $k=0.001 \impc$. When probing on this scale we lose the detection we had obtained at the optimized pivot. As this case shows, particularly in the presence of foreground uncertainties probing at the pivot scale, where the instrument is most sensitive, may constitute the difference between detection and non-detection of primordial tensors.

\subsection{Transforming between pivots}

We now compare the constraints obtained under different prior assumptions and at different pivot scales. We are particularly interested in studying the robustness of the posteriors on $\AT$ and $\nT$ in response to such changes. In Fig.~\ref{posteriorTransform} we compare the contours obtained under a linear prior on $\AT$, left panel, and a logarithmic prior on $\AT$, right panel. The linear prior has the range $0<10^{10} \AT < 100$ and the logarithmic prior  $-6 < \log 10^{10} \AT < 3$.
The blue contours in both panels are obtained at {\it Planck}+WP+BICEP2 pivot, $k=0.015 \impc$. We then repeat the procedure, taking a linear and logarithmic prior in $\AT$, but sampling instead at {\it Planck}+WP pivot scale, $k=0.001\impc$. Lastly, we take these posteriors at Planck's pivot and transport them to {\it Planck}+WP+BICEP2 pivot. These correspond to the red contours, linear on the left, and logarithmic on the right. We superpose these over the blue contours which are originally run at the {\it Planck}+WP+BICEP2 pivot. 

We are interested in identifying the prior which leaves the posterior unchanged under variations of scale, i.e.\ the prior for which the superposition of the red and blue contours is the most similar. From Fig.~\ref{posteriorTransform} it is clear that this is the case in the logarithmic prior in the right panel, while the linear prior in the left panel gives rise to quite different posteriors under transformation between scales.

Still, the logarithmic case shows a mismatch of the confidence contours at large values of $\nT$. The red contours, sampled at the {\it Planck}+WP scale, exclude values of $\nT\gtrsim3$, which does not happen in the sampling at {\it Planck}+WP+BICEP2 pivot (blue). This is not an artefact caused by the logarithmic prior but rather because of sampling $\AT$ away from optimal the pivot scale. At {\it Planck}+WP pivot (but still using the {\it Planck}+WP+BICEP2 data), very blue values of $\nT$ will require very small values of $\AT$ and will be cut off by the prior (as shown one the right-hand column of  Fig.~\ref{dustOptimPessim}. When the chain is transformed to the {\it Planck}+WP+BICEP2 pivot, the lack of chain elements with $\nT>3$ and $\log 10^{10} \AT <-6$ manifests as a cutoff for very blue values of $\nT$. The same argument holds for the linear prior case, with the addition that a linear prior on $\AT$ at $k=0.001\impc$ transforms non-linearly to the new pivot (as shown in Fig.~\ref{priorTransform}). The small prior volume occupied by the small $\AT$ and blue $\nT$ region favoured by the data at the new pivot leads to ill-matching posterior distributions.

As we argued in Section~\ref{s:prior}, in an era where tensor detection is a goal, it is essential to safely transport posteriors of any two experiments having different pivot scales, like the case of \textit{Planck} and BICEP2. Figure~\ref{posteriorTransform} shows clearly that a prior logarithmic on $\AT$ is the preferred prior to ensure a robust characterization of the tensor spectrum.

Finally, we note that while the above analysis implies that if the BICEP2 signal has a detectably-large primordial component then $n_{\rm T}>0$, short-scale observations such as Big Bang Nucleosynthesis impose a fairly strict upper limit from avoiding having too large a density in sub-horizon gravitational waves at key epochs \cite{WX1,Smith:2014kka,Kuroyanagi}. This limit is typically around 0.5 with some dependence on cosmological modelling assumptions, and also depending on the amplitude of perturbations. This could be imposed as an additional constraint, perhaps formulated as a prior cut across the \mbox{$A_{\rm T}$--$\nT$} parameter space, within our framework.

\section{Conclusions}

Motivated by the BICEP2 detection of large-angle B-mode polarization and its possible primordial origin, in this article we have advocated a principled approach to executing analyses that aim to demonstrate detection of tensors. We have argued that the tensor spectrum should be constrained directly, rather than via the tensor-to-scalar ratio, which enables a clean identification of the `pivot' scale at which the tensors are optimally constrained. Particularly while observational data leave open the possibility of a tensor spectral index far from zero, we have highlighted the importance of setting a well-considered prior on the tensor amplitude at the pivot scale, arguing that a uniform (linear) prior on the amplitude is typically inappropriate.

We then reanalysed the {\it Planck}+WP+BICEP2 data combination. We did this first under the assumption of BICEP2 being entirely primordial, in order to enable comparison of our results with previous ones which used less well-motivated priors and pivot scales. Our results, shown in Fig.~\ref{bicep2pivot}, indicate a strong detection of tensors under this assumption and affirm the strongly blue-tilted tensor spectrum required to match all these datasets, with $\nT = 1.8 \pm 0.6$. This blue tilt means that the tensor-to-scalar ratio, when expressed at the pivot scale, appears larger than in the BICEP2 article \cite{BICEP2} which effectively reported on a larger scale. Our determination of $\nT$ as being significantly blue agrees with previous articles, e.g.~Refs.~\cite{Gerbino:2014,Chang:2014loa}.

It now seems much more plausible that the BICEP2 signal is significantly, or entirely, non-primordial with a substantial component due to polarized dust emission. Mortonson and Seljak \cite{Mortonson:2014} and Flauger et al.~\cite{Flauger:2014}  showed that plausible modelling of the dust readily eliminates the primordial tensor detection, and {\it Planck} has confirmed that the likely level of dust is sufficient to do this \cite{planckdust}. For our analysis, rather than modelling uncertainties in the dust we anticipate a future era where the dust properties may be accurately pinned down, for instance by further {\it Planck} and BICEP/Keck Array observations, and study the impact on future searches for primordial tensors. We focus on two incarnations of the simple {\it Planck}-motivated dust model of Mortonson and Seljak, an `optimistic' one which leaves a significant part of the signal available to be ascribed to a cosmological origin, and a `pessimistic' one that more or less subsumes the BICEP2 signal. The former scenario is at the lower limit of the dust contribution inferred from {\it Planck} 353~GHz observations \cite{planckdust}. 

As expected, we find an increasing dust signal lowers both the amplitude and detection significance of the tensors, while simultaneously weakening the constraint on $\nT$. With the optimistic dust model, a detection somewhat over 95\% confidence remains, but the required $\nT$ remains entirely in the positive region that is forbidden to normal inflation models. We therefore conclude that if there were a dust contribution strong enough to make the tensor signal compatible with simple inflation models, it would also be strong enough to eliminate the significance of the detection. Put another way, if we believed previous observations from {\it Planck}+WP, combined with the assumption $\nT<0$ from simple inflation models, we would have to conclude that there could not be a primordial signal strong enough to be detected by BICEP2, whose signal would need an alternative explanation such as polarized dust.

\begin{acknowledgments}
M.C.\ was supported by EU FP7 grant PIIF-GA-2011-300606, A.R.L.\  by the Science and Technology Facilities Council [grant numbers ST/K006606/1 and ST/L000644/1], and D.P.\ by an Australian Research Council Future Fellowship [grant number FT130101086]. Research at Perimeter Institute is supported by the Government of Canada through Industry Canada and by the Province of Ontario through the Ministry of Research and Innovation. A.R.L.\ acknowledges hospitality of the Perimeter Institute during part of this work. D.P.\ acknowledges hospitality of the Royal Observatory Edinburgh for part of this work. We thank Ed Copeland, Antony Lewis, and John Peacock for discussions.
\end{acknowledgments}

\end{document}